# Anisotropy in c-oriented MgB$_2$ thin films grown by Pulsed Laser Deposition

C.Ferdeghini[a], V.Ferrando[a], G.Grassano[a], W.Ramadan[a], E.Bellingeri[a], V.Braccini[a], D.Marrè[a], M.Putti[a], P.Manfrinetti[b], A.Palenzona[b], F. Borgatti[c], R. Felici[c], L.Aruta[d]

[a]INFM, Dipartimento di Fisica, Via Dodecaneso 33, 16146 Genova, Italy

[b]INFM, Dipartimento di Chimica e Chimica Industriale, Via Dodecaneso 31, 16146 Genova, Italy

[c]INFM-Operative Group in Grenoble, c/o ESRF, BP 220, F-38043 Grenoble, France

[d]INFM, INFM-University of Roma "Tor Vergata" Dept. S.T.F.E.,Via di Tor Vergata 110, Roma, Italy

## Abstract

The electronic anisotropy in MgB$_2$, is still a not completely clear topic; high quality c-oriented films are suitable systems to investigate this property. In this work we present our results on MgB$_2$ superconducting thin films grown on MgO and sapphire substrates. The films are deposited in high vacuum, at room temperature, by Laser Ablation, starting from two different targets: pure Boron and stoichiometric MgB$_2$. In both cases, to obtain and crystallize the superconducting phase, an ex-situ annealing in magnesium vapor is needed. The films were characterized by Synchrotron radiation diffraction measurements; the films turned out to be strongly c-oriented, with the c-axis perpendicular to the film surface and an influence of the substrate on crystallographic parameters is observed. Resisivity measurements with the magnetic field perpendicular and parallel directions to the film surface evidenced an anisotropic upper critical field behavior. The $H_{c2}$ ratios ($\eta$) resulted in the range 1.2-1.8; this difference will be discussed also in comparison with the literature data.

**Corresponding author:**

C.Ferdeghini, INFM Dipartimento di Fisica, Via Dodecaneso 33, 16146 Genova, Italy.

Tel +39 0103536282, Fax +39 010311066, E-mail: Ferdeghini@fisica.unige.it

**Introduction**

The evaluation of the anisotropy in superconducting materials is very important both for basic understanding and for practical applications. In $MgB_2$ case this topic is still under debate; in literature, indeed, the reported anisotropy values η (defined as the ratio between the critical fields parallel and perpendicular to the basal planes) range between 1.1 and 9 [3-6]. In order to clearly evaluate the anisotropy in this compound, measurements should be performed on oriented samples as single crystals or epitaxial thin films. Being the single crystal synthesis problematic due to the $MgB_2$ phase diagram, the availability of epitaxial (or at least c-oriented) thin films appear to be a crucial goal. However, also the thin film deposition is quite difficult due to the Magnesium volatility and to Magnesium and Boron reactivity with oxygen. Different methods for superconducting $MgB_2$ film preparation have been reported in literature: nearly all of them are based on a two-step process [1]. The first step is the room temperature deposition of a non-superconducting precursor layer, which is then annealed in a second step to obtain a superconducting phase. The widely used precursor layers are amorphous Mg-B films deposited from stoichiometric targets or amorphous boron films. Commonly, annealing is performed ex-situ in magnesium atmosphere. For electronic applications, which require the integration of different materials as in devices or junctions, however, the annealing procedures present some drawbacks and a single step in-situ deposition of superconducting $MgB_2$ is necessary. Up to now, only one method to produce superconducting samples in one step has been reported in literature [2], though samples obtained in such a way have a critical temperature of only 25K and the process still needs to be optimized.

Anisotropy measurements on c-oriented thin films appeared recently in literature. Patnaik et al. [7] reported anisotropy measurements on films grown on $SrTiO_3(111)$ starting from stoichiometric target. They found $\eta$ in the range 1.8-2. M.H.Jung et al [8], instead, found a lower value $\eta=1.25$, on epitaxial film grown on $Al_2O_3$ r-cut starting from Boron precursor. Experiments performed on small single crystals, pretty recently available, gave values in the interval 2.6-3 [9-11]. From this brief overview it is apparent that the issue about precise value of $\eta$ is still open.

Here we present films grown by Pulsed Laser Deposition starting from both $MgB_2$ sintered target prepared by direct synthesis from the elements [12] and from Boron target obtained by pressing amorphous B powders. Both amorphous precursor layers were deposited in high vacuum condition, at room temperature. To crystallize the superconducting phase, we carried out an *ex-situ* annealing procedure in magnesium vapor. The samples were placed in a sealed tantalum tube with Mg lumps (approx 0.05 mg/cm$^3$), in Ar atmosphere, and then in an evacuated quartz tube and heated at T= 850∞C for 30 minutes followed by a rapid quenching to room temperature, according to the procedure reported in [12].

We used MgO and Sapphire as substrates for its thermal stability at the high temperatures used in the subsequent annealing process. We used the (100) and (111) MgO cuts and the r-cut for the Sapphire. The MgO(100) crystallographic orientation has a square surface symmetry; the (111) orientation a hexagonal one (with a lattice mismatch with $MgB_2$ less than 3%) and the r-cut Sapphire substrate a rhombic surface symmetry.

**Structural characterization by synchrotron radiation**

To investigate the structural properties of some of our samples, X-ray diffraction measurements were performed by synchrotron radiation. Thanks to the high intensity and collimation of the synchrotron X-ray beam, we could perform high resolution diffraction measurements. With the exchanged momentum either perpendicular (symmetrical configuration) or parallel (grazing incidence configuration) to the sample surface we can determine the out-of-plane orientation and the in-plane structural characteristics, respectively. The measurements were carried out at the ID32 beamline at the ESRF, where the incoming X-ray beam from two undulators was monochromatized with a Si(111) double crystal monochromator selecting an energy of 11 keV.

Typical $\vartheta$-$2\vartheta$ scans in symmetrical configuration are shown in fig.1 for two samples grown, starting from stoichiometric target, on r-cut Sapphire (lower panel) and MgO(111) (upper panel) substrates. Both the (001) and (002) reflections of the $MgB_2$ phase are observed in the scan of sample grown on Sapphire, which is a clear evidence of a strong c-axis orientation. Moreover an intense peak of the (101) reflection is also present in the scan, due to a not well oriented fraction of the sample. However, the sample grown on MgO substrate is much more disordered, showing in the symmetrical scan of lower panel of fig.1 different reflections belonging to several disordered grains. A better orientation along the c-axis for the first sample is confirmed by the rocking curves of the (002) reflections also reported for both samples in fig.1. In the case of the sample on the Sapphire substrate a FWHM (Full Width at Half maximum) of almost 1.3° is obtained by fitting the peak with a Lorentzian shape, which is lower than the FWHM of the sample grown on MgO (111) substrate of almost 2.0°.

The in-plane orientation of the films was investigated by grazing incidence measurements, with a constant angle of the X-rays with the surface of almost 0.5°. This low angle of incidence was slightly higher than the critical angle, as determined experimentally and it was selected in order to increase the signal to background ratio, thus enhancing the scattering contributions from the film relative to the substrate. In this configuration the scattering vector is approximately parallel to the surface and by scanning the vertical $2\vartheta$ angle the in-plane reflections were detected. Both the samples have not shown a single in-plane orientation, even if with some differences. Indeed, in the sample on Sapphire a strong azimuthal dependence was observed for the (110) reflection and in the sample on MgO a strong azimuthal dependence was observed for the (100) reflection. These results are consistent with two different in-plane preferential orientations for the two samples. Also the in-plane lattice parameters are different, as calculated from the (100) reflections for both samples and reported in Table 1, together with the c-axis value calculated by the (002) reflection measured in symmetrical configuration. In the case of the sample grown on Sapphire a good estimation of the c-axis value could not be obtained, even if a miscut of almost 0.7° reduced the contribution of the substrate and other measurements were performed maintaining the exit angle equal to the incident value minus 1° to further reduce that strong contribution.

From this characterization a first indication of the influence of the substrate on the growth mechanism of these samples can be obtained. The sample grown on MgO (111) is strained even if the in-plane lattice is still not well matched with the substrate. The film try to accomodate the in-plane lattice with exagonal face of the substrate reducing the in-plane lattice parameters, but the mismatch is not low enough and the strain relief leads to an increase of the structural disorder. However the sample grown on Sapphire is fully relaxed to the bulk values and it is better c-axis oriented.

**Anisotropy measurements**

In this paragraph we present anisotropy measurements on three different films (film 1, 2, 3) with $T_C$ = 31.5K, 36.2K and 37.5K respectively. The electrical film characteristics ($T_C$, $\Delta T_C$ and RRR) are reported in table 2. Film 2 is the film grown on $Al_2O_3$ r-cut from stoichiometric target whose structural characteristics are illustrated in the lower panel of fig.1. Film 1 is grown on MgO(100) from stoichiometric target and film 3 is deposited on MgO(111) from Boron target. To evaluate the structural properties of film1 and film3 we performed standard x-rays in the $\vartheta$-$2\vartheta$ Bragg-Brentano geometry. We define a texturing coefficient $t = \left[\frac{I^{film}(002)}{I^{film}(101)}\right]\left[\frac{I^{powder}(101)}{I^{powder}(002)}\right]$ where $I^{film}(002)$ and $I^{film}(101)$ are the measured intensities of the reflection (002) and (101) and $I^{powder}(002)$ and $I^{powder}(101)$ are the tabulated intensity of randomly oriented powders. In table 2 we report $t$=1.6, 22.6 and 5.6 for films 1, 2, 3 respectively thus concluding that the films are preferentially c-oriented and in particular, film 2 more than film 3 and film 3 more than film 1. The rocking curves around the (002) peak confirm this trend and FWHM values of about 8° for the film 1, 2° for the film 2 and about 3° for film 3 are obtained. All these data are summarized in table 2.

Electrical resistance measurements as a function of temperature in applied magnetic field up to 9T were performed in a Quantum Design PPMS apparatus by using a four-probe AC resistance technique at 7 Hz. The magnetic field was applied parallel and perpendicular to the film surface and the current was always perpendicular to the magnetic field. $H_{c2}$ vs. temperature curves were determined for each field from magnetoresistivity measurements at the point of the transition in which the resistance is the 90% of the normal state value. The $H_{c2}$ evaluation is strongly criterion dependent mainly in samples with large inhomogeneity as film 3 ($\Delta T_c \sim 6$). In figure 2 we report $H_{c2}$ as a function of temperature for the three films in the two orientations: $H_{c2}$ are considerably higher when the field is parallel to the film surface.

We can observe, for film1 ( low $T_C$ and RRR values), a linear $H_{c2}$ versus T dependence near $T_C$, while for film3 ($T_C$ close to the bulk value and higher RRR) we observe the upward curvature typical of the clean limit. Film 2 shows a less pronounced upward curvature suggesting a crossover between dirty and clean limit in agreement with intermediate $T_c$ and RRR values.

From $H_{c2}$ (0) values, extrapolated with a BCS formula, it is possible to calculate the anisotropy factor η that resulted to be 1.8 for film 1, 1.2 for film 2 and 1.4 for film 3. These values are considerably lower in respect to the single crystal ones and this could derive from the not complete film orientation. Nevertheless we must consider that film 2 and 3 resulted to be more oriented than film 1 but they present lower η; therefore the difference in η among the three films cannot be related with the different grade of texturing. In any case these values lie in the literature range [7,8] despite the different substrates used. We believe that when $T_C$ increases also the RRR factor increases indicating a passage between dirty and clean limit conditions. In dirty limit we observe high anisotropy values, nearly temperature independent, and in clean limit a low (at least near $T_c$) temperature dependent anisotropy. The boron precursor method more easily produces samples with higher $T_C$ and RRR values and therefore clean limit conditions.

**References**

[1] C.Buzea, T.Yamashita, Cond-Mat/0108265

[2] G.Grassano et al, Supercond. Sci. Technol. 14 (2001) 1

[3] O.F. De Lima et al, Cond-mat/0103287.

[4] A.Andstein et al Cond-mat/0103408.

[5] F. Simon et al, Cond-mat/0104557.

[6] S.L.Budíko et al Cond-mat/0106577.

[7] S. Patnaik et al., Supercond. Sci. Technol. 14 (2001) 315

[8] M.H. Jung, Cond-mat/ 0106146.

[9] M. Xu et al, Cond-mat/0105271.

[10] S. Lee et al, Cond-mat/ 0105545.

[11] Kijoon H. P. Kim et al, Cond-mat/0105330.

[12] C.Ferdeghini et al., Cond-mat/0107031.

Table1. Lattice parameters of the two films of fig.1

| bulk | Al$_2$O$_3$ r-cut | MgO (111) |
|---|---|---|
| a=3.086 ≈ | a=3.083 ≈ | a=3.073 ≈ |
| c=3.520 ≈ | c= - | c=3.513 ≈ |

Table 2. Characteristics of the three films for anisotropy measurements.

| Sample | Film1 | Film2 | Film3 |
|---|---|---|---|
| Preparation | From $MgB_2$ | From $MgB_2$ | From Boron |
| Substrate | MgO(100) | $Al_2O_3$ r-cut | MgO(111) |
| $T_c$, K | 31.4 | 36.2 | 37.5 |
| $\Delta T_c$, K | 1.1 | ~6 | 0.6 |
| RRR | ~1 | 1.2 | 2.4 |
| Texturing | 1.6 | 22.6 | 5.6 |
| FWHM(002), degree | ~8 | 2 | 3 |
| $\eta$ | 1.8 | 1.2 | 1.4 |

**Figures captions**

Fig.1 $\vartheta$-$2\vartheta$ scans in symmetrical configuration and rocking curves of the (002) reflections for two samples grown, starting from stoichiometric target, on r-cut Sapphire (lower panel) and MgO(111) (upper panel) substrates.

Fig.2 $H_{c2}$ versus temperature for film1, film 2 and film 3 for magnetic fields parallel (full symbols) and perpendicular (open symbols) to the film surface.

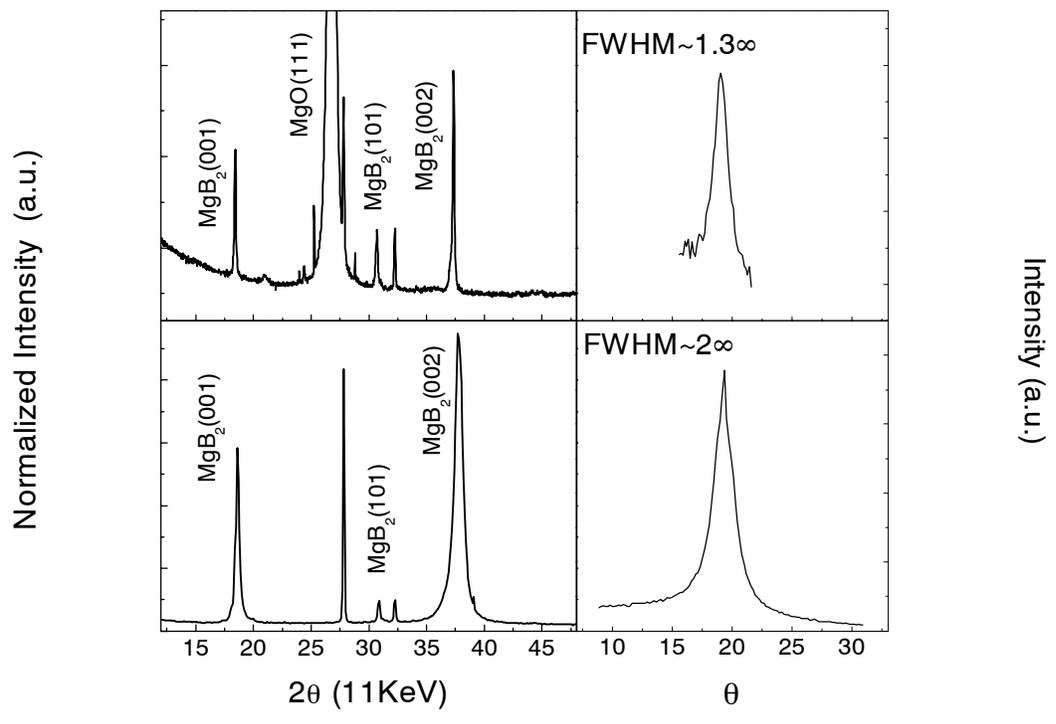

Figure 1

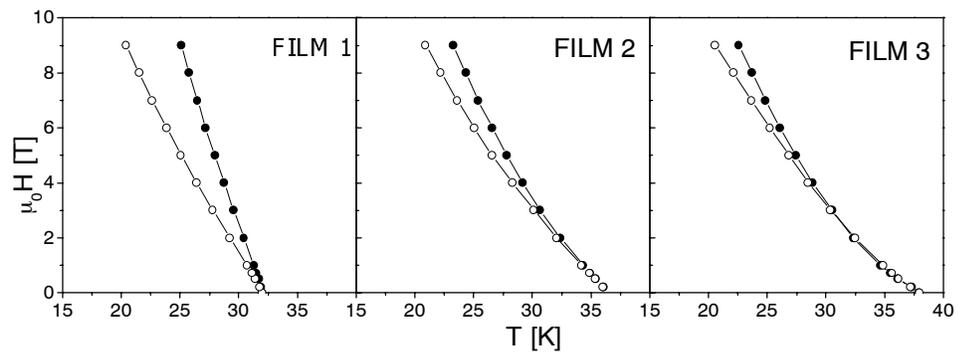

Figure 2